\documentclass[conference]{IEEEtran}
\usepackage[T1]{fontenc}
\usepackage{type1cm} 
\usepackage[cmex10]{amsmath}
\usepackage{amsthm} \theoremstyle{definition}
\usepackage[varg]{txfonts}
\usepackage[normalem]{ulem}
\usepackage[nospace]{cite}

\def\Eq#1{Eq.\@\,(\ref{#1})}
\def\Eqs#1{Eqs.\@\,(\ref{#1})}
\def\Thm#1{Theorem~\ref{#1}}
\def\Lma#1{Lemma~\ref{#1}}
\def\Sect#1{Sect.\@\,\ref{#1}}

\def\Prop#1{Prop.\@\,\ref{#1}}
\def\Coro#1{Corol.\@\,\ref{#1}}
\def\Def#1{Def.\@\,\ref{#1}}

\def\ie{{i.\@e.\@,~}}
\def\eg{{e.\@g.\@,~}}

\def\rank{\mathsf{rank\,}}
\def\dim{\mathsf{dim\,}}

\def\F{\mathbb{F}}

\def\c{\mathcal{C}}
\def\dc{\mathcal{C}^{\perp}}
\def\d{\mathcal{D}}
\def\dd{\mathcal{D}^{\perp}}

\def\colinv{\Gamma(\F_{q^m}^n)}
\def\colinvi#1{\Gamma_{#1}(\F_{q^m}^n)}
\def\equivocation{{\theta}}
\def\uequivocation{{\Theta}}
\def\ustrong{{\Omega}}

\def\bari#1{\,\overline{{\!\{#1\}\!}}\,}
\def\vecxi{Z_{\bari{i}}}

\newtheorem{theorem}{Theorem}
\newtheorem{lemma}[theorem]{Lemma}
\newtheorem{proposition}[theorem]{Proposition}
\newtheorem{corollary}[theorem]{Corollary}

\newtheorem{definition}[theorem]{Definition}
\newtheorem{example}[theorem]{Example}

\def\spf{\hspace{0.551ex}}
\begin{document}
\title{New{\spf}Parameters{\spf}of{\spf}Linear{\spf}Codes{\spf}Expressing{\spf}Security{\spf}Performance{\spf}of{\spf}Universal{\spf}Secure{\spf}Network{\spf}Coding}
\author{
\IEEEauthorblockN{
Jun KURIHARA\IEEEauthorrefmark{1}\IEEEauthorrefmark{2},
 Tomohiko UYEMATSU\IEEEauthorrefmark{1}
 and Ryutaroh MATSUMOTO\IEEEauthorrefmark{1}
}
\IEEEauthorblockA{
\IEEEauthorrefmark{1}Department of Communications and Integrated Systems, Tokyo Institute of Technology\\
2--12--1 Ookayama, Meguro-ku, Tokyo, 152--8550 Japan\\
Email: kurihara@kddilabs.jp, uyematsu@ieee.org, ryutaroh@rmatsumoto.org
}
\IEEEauthorblockA{
\IEEEauthorrefmark{2}KDDI R\&D Laboratories, Inc.\\
2--1--15 Ohara, Fujimino-shi, Saitama, 356--8502 Japan}
}
\maketitle
\begin{abstract}
The universal secure network coding presented by Silva et al.\@ 
realizes secure and reliable transmission of a secret message over any 
underlying network code, by using maximum rank distance codes.
Inspired by their result,
this paper considers the secure network coding based on arbitrary
linear codes, and investigates its security performance
and error correction capability that are guaranteed
independently of the underlying network code.
The security performance and error correction capability are
said to be \textit{universal}
when they are independent of underlying network codes.
This paper introduces new code parameters,
the relative dimension/intersection profile (RDIP) and the
relative generalized rank weight (RGRW) of linear codes.
We reveal that the universal security performance
 and universal error correction capability of secure network coding
 are expressed in terms of the RDIP and RGRW of linear codes.
The security and error correction of existing schemes 
are also analyzed as applications of the RDIP and RGRW.
\end{abstract}

\section{Introduction}\label{sect:intro}
In the scenario of \textit{secure network coding} introduced by Cai et al.\@ \cite{Cai2011},
a source node transmits $n$ packets from $n$ outgoing links to sink
nodes through a network that
implements network coding \cite{Ahlswede2000a,Koetter2003,Li2003},
and each sink node receives $n$ packets from $n$ incoming links.
In the network, there is a wiretapper who observes $\mu (< n)$ links.
The problem is how to encode a secret message into $n$ transmitted
packets at the source node, in such a way that the wiretapper obtain no
information about the message in the sense of information theoretic security.

As shown in \cite{ElRouayheb2012}, secure network coding can be seen as
a generalization of the wiretap channel II \cite{Ozarow1984} or secret
sharing schemes based on linear codes \cite{Chen2007,Duursma2010} for
network coding.
Hence, in secure network coding, the secrecy is realized by introducing
the randomness into $n$ transmitted packets as follows.
Suppose the message is represented by $l$ packets
$S_1,\dots,S_l$ $(1 \leq l \leq n)$.
Then, the source node encodes $(S_1,\dots,S_l)$ together with $n-l$ random
packets by linear codes,
and generates $n$ transmitted packets 
\cite{Silva2011,Ngai2011,ElRouayheb2012}.

Silva et al. \cite{Silva2011} proposed the \textit{universal secure
network coding} that is based on maximum rank distance (MRD) codes \cite{Gabidulin1985}.
Their scheme was universal in the sense that their scheme guarantees that over \textit{any}
 underlying network code, no information about $S$ leaks out even if
any $n-l$ links are observed by a wiretapper.
As shown in \cite{Silva2011}, their scheme with MRD codes is optimal in
terms of security and communication rate.
However, there exists some restrictions in universal secure network
coding with MRD codes.
In their scheme, the network must transport packets of size $m \geq n$.
The MRD code used in the scheme is defined over an $\F_{q^m}^n$,
where $\F_{q^m}$ is an $m$-degree field extension of a field $\F_q$ with
order $q$.
Thus, the size of the field $\F_{q^m}$ increases exponentially with $m$,
and the restriction of MRD codes with $m \geq n$ invokes the large
computational cost for encoding and decoding of MRD codes if $n$ is
large.
It is undesirable especially
in resource constraint environments.

Considering secure network coding without such a restriction,
Ngai et al.\@ \cite{Ngai2011}, and later Zhang et al.\@ \cite{Zhang2009},
investigated the security performance of secure network coding based on
general linear codes.
They introduced a new parameter of linear codes,
called the \textit{relative network generalized Hamming weight} (RNGHW),
and revealed that the security performance is expressed in terms of the
RNGHW.
The RNGHW depends on the set of coding vectors of
the underlying network code.
Hence, the RNGHW is not universal.

The aim of this paper is to investigate the security performance of
universal secure network coding based on general linear codes,
which is always guaranteed over \textit{any} underlying network code,
even over random network code.
This paper defines the universal security performance by the following two criteria.
One is called the \textit{universal equivocation} $\uequivocation_\mu$
that is the minimum uncertainty of the message under observation
of $\mu (< n)$ links, guaranteed independently of the underlying network code.
The other is called the \textit{universal $\ustrong$-strong security},
where $\ustrong$ is a performance measure such that no part of the secret
message is deterministically revealed even if at most $\ustrong$ links are observed.
The paper \cite{Kurihara2012} proposed a specific construction of the secure network
coding that attains the universal $(n-1)$-strong security,
and such a scheme is called universal strongly secure network coding \cite{Silva2009}.
Namely, the definition of universal $\ustrong$-strong security given in
this paper is a generalization of
universal strongly secure network coding considered in
\cite{Kurihara2012,Silva2009} for the number of tapped links.

In order to express $\uequivocation_\mu$ and $\ustrong$ in terms of code parameters,
this paper introduces two parameters of linear codes,
 called the \textit{relative dimension/intersection profile} (RDIP)
and the \textit{relative generalized rank weight} (RGRW).
The RGRW is a generalization of the minimum rank distance
\cite{Gabidulin1985} of a code.
We reveal that $\uequivocation_\mu$
and $\ustrong$ can be expressed in terms of the RDIP and the RGRW of the
codes.
Duursma et al.\@ \cite{Duursma2010}
first observed that the \textit{relative generalized Hamming weight} \cite{Luo2005}
exactly expresses the security performance and the
error correction capability of secret sharing.
Our definitions of RGRW and RDIP are motivated by their
result \cite{Duursma2010}.

Assume that the attacker is able not only to eavesdrop
 but also to inject
 erroneous packets anywhere in the network.
Also assume that
 the network may suffer from the rank deficiency
of the transfer matrix at a sink node.
Silva et al.\@'s scheme based on MRD codes \cite{Silva2011} enables to
correct such errors and rank deficiency at each sink node,
where its error correction capability is guaranteed over any
underlying network code, \ie universal.
This paper also generalizes their result and reveals
that the universal error correction capability of secure network
coding based on arbitrary linear codes can be expressed in
terms of the RGRW of the codes.

The remainder of this paper is organized as follows.
\Sect{sect:prelimi} presents basic notations, and introduces linear network coding.
\Sect{sect:wiretap} defines the universal security performance and universal error correction
 capability of secure network coding over wiretap network.
\Sect{sect:defrdiprgrw} defines the RDIP and RGRW of linear codes, and introduces their basic properties.
In \Sect{sect:universalsecure}, the universal security performance is
 expressed in
terms of the RDIP and RGRW.
The security of existing schemes \cite{Kurihara2012,Silva2009,Silva2011}
is also analyzed as applications of the RDIP and RGRW
in Examples \ref{ex1} and \ref{ex2}.
\Sect{sect:errorcorrection} gives the expression of the
universal error correction capability in terms of the RGRW, and also
analyze the error correction of \cite{Silva2011} by the RGRW in Example \ref{ex3}.

\section{Preliminary}\label{sect:prelimi}
\subsection{Basic Notations}
Let $H(X)$ be the Shannon entropy for a random variable $X$,
$H(X|Y)$ be the conditional entropy of $X$ given $Y$,
 and $I(X;Y)$ be the mutual information between $X$ and $Y$
 \cite{Cover2006}.
We write $|\mathcal{X}|$ as the cardinality of a set $\mathcal{X}$.
The entropy and the mutual information are
always computed by using $\log_{q^m}$.

Let $\F_q$ stand for a finite field containing $q$ elements
and $\F_{q^m}$ be an $m$-degree field extension of $\mathbb{F}_q$
 ($m \geq 1$).
Let $\F_q^n$ denote an $n$-dimensional row vector space over $\F_q$.
Similarly, $\F_{q^m}^n$ stands for an $n$-dimensional row vector space
over $\F_{q^m}$.
Unless otherwise stated,
we consider subspaces, ranks, dimensions, etc, over the field extension $\F_{q^m}$
instead of the base field $\F_q$.

An $[n,k]$ linear code $\c$ over $\F_{q^m}^n$
is a $k$-dimensional subspace of $\F_{q^m}^n$.
Let $\dc$ denote a \textit{dual code} of a code $\c$.
A subspace of a code is called a \textit{subcode} \cite{MacWilliams1977}.
For $\c \subseteq \F_{q^m}^n$,
we denote by $\c|\F_{q}$ a \textit{subfield subcode} of $\c$ over $\F_q$
\cite{MacWilliams1977}.
Observe that $\dim \c$ means the dimension of $\c$ as a vector space
over $\F_{q^m}$ whereas $\dim \c|\F_q$ is the dimension of $\c|\F_q$
over $\F_q$.

For a vector $\vec{v}=[v_1,\dots,v_n]\in\F_{q^m}^n$
and a subspace $V \subseteq \F_{q^m}^n$,
we denote $\vec{v}^{\,q} = [v_1^q,\dots,v_n^q]$
and $V^q = \{\vec{v}^{\,q} : \vec{v} \in V\}$.
Define a family of subspaces $V\subseteq\F_{q^m}^n$ satisfying $V = V^q$
by
$\colinv \triangleq \{\text{subspace } V \subseteq \F_{q^m}^n : V = V^q\}$.
Also define $\colinvi{i} \triangleq \{V\in\colinv : \dim V = i\}$.
For a subspace $V \!\subseteq\! \F_{q^m}^n$,
the followings are equivalent:
1) $V \!\in\! \colinv$;
2) $\dim V \!=\! \dim V|\F_q$ \cite[Lemma 1]{Stichtenoth1990}.

\subsection{Linear Network Coding}\label{sect:linearnetwork}
As in \cite{Silva2011,Ngai2011,Zhang2009,Cai2011,ElRouayheb2012},
we consider a multicast communication network represented by a directed
multigraph with unit capacity links, a single source node, and multiple
sink nodes.
We assume that \textit{linear network coding} \cite{Li2003,Koetter2003} is
employed over the network.
Elements of a column vector space $\F_q^{m \times 1}$ are called \textit{packets}.
Assume that each link in the network can carry a single $\F_q$-symbol per
one time slot, and that each link transports a single packet over $m$ time
slots without delays, erasures, or errors.

The source node produces $n$ packets
$X_1$, \ldots, $X_n\in \F_q^{m \times 1}$
and transmits $X_1$, \ldots, $X_n$ on $n$ outgoing links
over $m$ consecutive time slots.
Define the $m \times n$ matrix $X=[X_1,\dots,X_n]$.
The data flow on any link can be represented as an $\F_q$-linear
combination of packets $X_1,\dots,X_n \in \F_q^{m \times 1}$.
Namely, the information transmitted on a link $e$ can be denoted as
 $b_e X^{\rm T} \in \F_q^{1 \times m}$,
 where $b_e \in \mathbb{F}_q^n$ is called
 a \textit{global coding vector} (GCV) of $e$.
Suppose that a sink node has $N$ incoming links.
Then, the information received at a sink node can be represented as an
 $N \times m$ matrix $AX^{\rm T} \in\F_q^{N \times m}$,
 where $A\in\F_q^{N \times n}$ is the transfer matrix
 constructed by gathering the GCV's of $N$ incoming links.
The network code is called \textit{feasible} if every transfer
matrix to a sink node has rank $n$ over $\F_q$.
The system is called \textit{coherent} if $A$ is known to each
sink node; otherwise, called \textit{noncoherent}.

\section{Universal Security Performance and Universal Error Correction Capability of Secure Network Coding}\label{sect:wiretap}
This section introduces the wiretap network model with packet errors
and the nested coset coding scheme in secure network coding
\cite{ElRouayheb2012,Silva2011,Zhang2009,Ngai2011}.
Then, we define the universal security performance
in terms of the \textit{universal equivocation} and
the \textit{universal $\ustrong$-strong security} on the wiretap
network model.
We also define the universal error correction capability of secure network coding.
From now on, only one sink node is assumed without loss of generality.
In addition, we focus on the fundamental case of coherent systems
in this paper due to the space constraint.
But, as in \cite{Silva2011},
all analysis in this paper can be easily adapted to the
case of noncoherent systems.

\subsection{Wiretap Networks with Errors, and Nested Coset Coding}\label{sect:nestedcoding}

Following \cite{Cai2011,Silva2011,Ngai2011,Zhang2009,ElRouayheb2012},
assume that in the setup of \Sect{sect:linearnetwork},
there is a wiretapper who has access to packets transmitted on any $\mu$ links.
Let $\mathcal{W}$ be the set of $|\mathcal{W}|=\mu$ links observed by the wiretapper.
Then the packets observed by the wiretapper are given by
$W^{\rm T} = B_\mathcal{W} X^{\rm T}$, where rows of $B_\mathcal{W} \in \F_q^{\mu \times n}$
are the GCV's associated with the links in $\mathcal{W}$.

In the scenario \cite{ElRouayheb2012,Silva2011,Zhang2009,Ngai2011}, the
source node first regards an $m$-dimensional column vector space
$\F_q^{m \times 1}$ as $\F_{q^m}$,
and fix $l$ for $1 \!\leq\! l \!\leq\! n$.
Let $S\!=\![S_1,\dots,S_l] \!\in\! \F_{q^m}^l$
be the secret message, and assume that $S_1,\dots,S_l$ are uniformly
distributed over $\F_{q^m}^l$ and mutually independent.
Under the wiretapper's observation,
the source node wants to transmit $S$ without information leakage to
the wiretapper.
To protect $S$ from the wiretapper,
the source node encodes $S$ to a transmitted vector
$X\!=\![X_1,\dots,X_n]\!\in\!\F_{q^m}^n$ of $n$ packets by
 applying the \textit{nested coset coding scheme}
 \cite{Zamir2002,Subramanian2009,Chen2007,Duursma2010}
 on $S$.
In \cite{Duursma2010,Chen2007}, its special case is called a \textit{secret
sharing scheme based on linear codes}.
\begin{definition}[Nested Coset Coding Scheme]\label{def:nestedcoding}
Let $\c_1 \subseteq \F_{q^m}^n$ be a linear code over $\F_{q^m}$ ($m \geq 1$),
and $\c_2 \subsetneqq \c_1$ be its subcode with dimension $\dim \c_2 = \dim \c_1 - l$ over $\F_{q^m}$.
Let $\psi:\F_{q^m}^l\rightarrow\c_1/\c_2$ be an arbitrary isomorphism.
For a secret message $S\in\F_{q^m}^l$,
we choose $X$ from a coset $\psi(S) \in \c_1/\c_2$
 uniformly at random and independently of $S$.
\end{definition}
Then, the source node finally transmit $X$ over the network coded
network.
\Def{def:nestedcoding} includes the Ozarow-Wyner coset coding scheme \cite{Ozarow1984}
as a special case with $\c_1=\F_{q^m}^n$.
Hence, when we set $\c_1=\F_{q^m}^n$,
this is the secure network coding based on Ozarow-Wyner
coset coding scheme \cite{Ngai2011,Silva2011,ElRouayheb2012}.

Corresponding to $X$ transmitted from the source node,
 the sink node receives a vector of $N$ packets
$Y \in \F_{q^m}^N$.
Here we extend the basic network model described in
\Sect{sect:linearnetwork} to incorporate packet errors and rank
deficiency of the transfer matrix $A \in \F_q^{N \times n}$ of the sink node.
Suppose that at most $t$ errors can occur in any of links, causing
the corresponding packets to become corrupted.
Then, as \cite{Silva2009a},
$Y$ can be expressed by
\begin{align*}
Y^{\rm T}=AX^{\rm T}+DZ^{\rm T},
\end{align*}
where $Z\in\F_{q^m}^t$ is the $t$ error packets,
 and $D\in\F_q^{N \times t}$ is the transfer matrix of $Z$.
We define $\rho \triangleq n-\rank A$ as the rank deficiency of $A$.
In this setup, we want to decode $S$ correctly from $Y$.
If the network is free of errors and the network code used is feasible,
$X$ can be always reconstructed from $Y^{\rm T}=AX^{\rm T}$ as
 described in \Sect{sect:linearnetwork}.
Then, the coset $\psi(S)$, and hence $S$, is uniquely determined from $X$
from \Def{def:nestedcoding}.

\subsection{Definition of Universal Security Performance}\label{sect:securityperformance}
The security performance of secure network coding in the
above model was measured by the following criterion \cite{Zhang2009,Ngai2011}.
\begin{definition}[Equivocation]\label{def:nonuniversalperformance}
The minimum uncertainty $\equivocation_\mu$ of $S$
 given $B_\mathcal{W}X^{\rm T}$ for all possible $\mathcal{W}$'s ($|\mathcal{W}|=\mu$) in
 the network is called \textit{equivocation},
defined as
${\displaystyle
\equivocation_\mu \!\triangleq\!
 \min_{\mathcal{W}: |\mathcal{W}|=\mu}
 H(S|B_\mathcal{W}X^{\rm T})}$.
\end{definition}
As defined in \Def{def:nonuniversalperformance},
$\equivocation_\mu$ depends on the underlying network code.
In \cite{Ngai2011,Zhang2009},
$\equivocation_\mu$ for $m=1$
was expressed in terms of the relative network generalized Hamming weight
(RNGHW) of $\c_1$ and $\c_2$.
The RNGHW is the value determined according to GCV's of all
links in the network.
Hence, the RNGHW cannot determine the equivocation over
random linear network code \cite{Ho2006}.
Here, we extend \Def{def:nonuniversalperformance} by
requiring the independence of the underlying network code, as follows.
\begin{definition}[Universal Equivocation]\label{def:universalperformance}
The \textit{universal equivocation} $\uequivocation_\mu$
is the minimum uncertainty of $S$
 given $BX^{\rm T}$ for all $B\in\F_q^{\mu \times n}$, defined as
\begin{align*}
\uequivocation_{\mu} \triangleq \min_{B \in \F_q^{\mu \times n}}H(S|BX^{\rm T}).
\end{align*}
\end{definition}
As defined in \Def{def:universalperformance},
$\uequivocation_\mu$ does not depend on the set of $\mathcal{W}$'s in
the network.
Silva et al.\@'s universal secure network coding scheme based on MRD
codes \cite{Silva2011} achieves
$\uequivocation_{n-l} = H(S)$ in \Def{def:universalperformance} provided $m \geq n$.

\Def{def:universalperformance} defines the security for the whole
components of a message $S=[S_1,\dots,S_l]$.
Here we focus on the security for every part of $S$,
and give the following definition.
\begin{definition}[Universal $\ustrong$-Strong Security]\label{def:universalalpha}
Let $S_\mathcal{Z}=(S_i:i\in\mathcal{Z})$ be a tuple for a subset
 $\mathcal{Z}\subseteq\{1,\dots,l\}$.
We say that a secure network coding scheme attains the \textit{universal
 $\ustrong$-strong security} if we have
\begin{align}
I(S_\mathcal{Z};BX^{\rm T})&=0,\quad
 \forall \mathcal{Z},
 \forall B \in \F_q^{(\ustrong-|\mathcal{Z}|+1) \times n}.\label{eq:universalalpha}
\end{align}
\end{definition}
As \cite{Harada2008,Matsumoto2011,Silva2009},
a scheme with universal $\ustrong$-strong security does not leak
any $|\mathcal{Z}|$ components of $S$ even if at most $\ustrong-|\mathcal{Z}|+1$ links are
observed by the wiretapper.
Moreover, this guarantee holds over any underlying
network code as $\uequivocation_\mu$.
We note that if a scheme achieves the $\ustrong$-strong security,
the universal equivocation $\uequivocation_\mu$ 
 for $\mu=\ustrong-l+1$
must be $\uequivocation_{\ustrong -l+1}=H(S)$ as shown in \Def{def:universalalpha}.
However, the converse does not always hold.

The scheme in \cite{Kurihara2012} achieves $\ustrong=n-1$ provided $m \geq l+n$
by nested coset coding with MRD codes.
The universal strongly security in \cite{Silva2009}
is a special case of \Def{def:universalalpha} with
$\ustrong =n-1$.
\subsection{Definition of the Universal Error Correction Capability of Secure Network Coding}\label{sect:erroneousnetwork}
In the model described in \Sect{sect:nestedcoding},
 the error correction capability of secure network coding, guaranteed
 over any underlying network code, is defined as follows.
\begin{definition}[Universally $t$-Error-$\rho$-Erasure-Correcting
 Secure Network Coding]
A secure network coding scheme is called \textit{universally
 $t$-error-$\rho$-erasure-correcting}, if
\begin{align*}
&H(S|Y)=0,\quad Y^{\rm T}=AX^{\rm T}+DZ^{\rm T},\\
&\quad
\forall A \!\in\!\F_q^{N \times n}: \rank A\!\geq\!n\!-\!\rho,
\forall X \in \psi(S),
\forall D \!\in\!\F_q^{N \times t},
\forall Z \!\in\! \F_{q^m}^t,
\end{align*}
\ie $S$ can be uniquely determined from $Y$ against
$t$ errors over any underlying network code with at most
 $\rho$ rank deficiency.
\end{definition}
Silva et al.\@'s scheme \cite[Section VI]{Silva2011} is
universally $t$-error-$\rho$-erasure-correcting when the minimum rank distance
\cite{Gabidulin1985} of $\c_1$ is greater than $2t+\rho$.

\section{New Parameters of Linear Codes and Their Properties}\label{sect:defrdiprgrw}
This section introduce
 the \textit{relative dimension/intersection profile} (RDIP) and the
 \textit{relative generalized rank weight} (RGRW) of linear codes.
In the following sections, these parameters are used to characterize
 the universal security performance and the universal error correction
 capability of secure network coding.
\subsection{Definition}
We first define the \textit{relative dimension/intersection
profile} (RDIP) of linear codes as follows.
\begin{definition}[Relative Dimension/Intersection Profile]\label{def:rdip}
Let $\c_1 \subseteq \F_{q^m}^n$ be a linear code and $\c_2\subsetneqq\c_1$ be its
 subcode.
Then, the $i$-th relative dimension/intersection profile (RDIP) of $\c_1$ and $\c_2$ is
the greatest difference between dimensions over $\F_{q^m}$ of
intersections, defined as
\begin{align}
K_{R,i} (\c_1,\c_2) \triangleq
\max_{V \in \colinvi{i}}
\left\{
\dim(\c_1 \cap V) - \dim(\c_2 \cap V)
\right\}, \label{eq:defrdip}
\end{align}
for $0 \leq i \leq n$.
\end{definition}
Next, we define the \textit{relative generalized rank weight} (RGRW) of
linear codes as follows.
\begin{definition}[Relative Generalized Rank Weight]\label{def:rgrw}
Let $\c_1 \subseteq \F_{q^m}^n$ be a linear code and $\c_2\subsetneqq\c_1$ be its
 subcode.
Then, the $i$-th relative generalized rank weight (RGRW) of $\c_1$ and
 $\c_2$ is defined by
\begin{align}
&M_{R,i}(\c_1,\c_2) \nonumber\\
& \!\triangleq\!
\min
\left\{ \dim V : 
V \!\in\! \colinv,
\dim(\c_1 \!\cap\! V) \!-\! \dim(\c_2 \!\cap\! V) \!\geq\! i
\right\},\label{eq:defrgrw}
\end{align}
for $0 \leq i \leq \dim (\c_1/\c_2)$.
\end{definition}
The relative dimension/length profile and the relative
generalized Hamming weight introduced in \cite{Luo2005}
are equivalent to \Eqs{eq:defrdip} and (\ref{eq:defrgrw})
with $\colinvi{i}$ and $\colinv$ replaced by suitable smaller
sets, respectively.

\subsection{Basic Properties of the RDIP and the RGRW, and the Relation between the Rank Distance and the RGRW}
This subsection introduces some basic properties of the RDIP and the RGRW,
 and also shows the relation between the RGRW and the rank distance \cite{Gabidulin1985}.
These will be used for expressions of the universal security performance
 and the universal error correction capability of secure network coding.

First, we introduce the following theorem and lemma about the RDIP and
the RGRW.

\begin{theorem}[Monotonicity of the RDIP]\label{thm:monotonerdip}
Let $\c_1 \subseteq \F_{q^m}^n$ be a linear code and $\c_2 \subsetneqq \c_1$ be
 its subcode.
Then, the $i$-th RDIP $K_{R,i}(\c_1,\c_2)$ is nondecreasing with $i$
 from $K_{R,0}(\c_1,\c_2)=0$ to $K_{R,n}(\c_1,\c_2)=\dim(\c_1/\c_2)$,
and
$0 \leq K_{R,i+1}(\c_1,\c_2)-K_{R,i}(\c_1,\c_2)\leq 1$ holds.
\end{theorem}
\begin{IEEEproof}
$K_{R,0}(\c_1,\c_2)=0$ and $K_{R,n}(\c_1,\c_2)=\dim(\c_1/\c_2)$,
 are obvious from \Def{def:rdip}.
Recall that
\begin{align*}
\colinvi{i} = \left\{V \subseteq \F_{q^m}^n :
V=\{\vec{u}G: \vec{u}\in\F_{q^m}^i\}, G\in\F_q^{i \times n}, \rank G = i
 \right\},
\end{align*}
for $1 \leq i \leq n$ from \cite[Lemma 1]{Stichtenoth1990}.
This implies that
for any subspace $V_1 \in \colinvi{i+1}$,
there always exist some $V_2$'s satisfying $V_2 \in \colinvi{i}$ and
$V_2 \subsetneqq V_1$.
This yields $K_{R,i}(\c_1,\c_2)\leq K_{R,i+1}(\c_1,\c_2)$.

Next we show that the increment at each step is at most $1$.
Consider arbitrary subspaces $V, V' \in \colinv$
 such that $\dim V'=\dim V + 1$ and $V \subsetneqq V'$.
Let
$f = \dim(\c_1 \cap V) - \dim(\c_2 \cap V)$;
$g = \dim(\c_1 \cap V') - \dim(\c_2 \cap V')$.
Since $\dim (\c_1 \cap V) +1 \!\geq\! \dim (\c_1 \cap V') \!\geq\! \dim (\c_1 \cap V)$
and $\c_2 \subsetneqq \c_1$,
we have $f+1 \geq g \geq f$ and hence
$K_{R,i}(\c_1,\c_2)+1 \geq K_{R,i+1}(\c_1,\c_2) \geq K_{R,i}(\c_1,\c_2)$.
\end{IEEEproof}
\begin{lemma}\label{lma:rgrw}
Let $\c_1 \subseteq \F_{q^m}^n$ be a linear code and $\c_2 \subsetneqq \c_1$
 be its subcode.
Then, the $i$-th RGRW $M_{R,i}(\c_1,\c_2)$ is strictly increasing with $i$.
Moreover, $M_{R,0}(\c_1,\c_2)=0$ and
\begin{align*}
&M_{R,i}(\c_1,\c_2)
 =
\min
\left\{ j : 
K_{R,j}(\c_1,\c_2)
 = i
\right\}\\
 &\ =
\min
\left\{ \dim V : V \in \colinv,
\dim(\c_1 \cap V) - \dim(\c_2 \cap V)
 = i
\right\},
\end{align*}
where $0\leq i \leq \dim (\c_1/\c_2)$.
\end{lemma}
\begin{IEEEproof}
First we have
\begin{align*}
&\min\left\{ j : 
K_{R,j}(\c_1,\c_2)
 \geq i
\right\}\\
&\!=\!
\min
\big\{
j : \exists V \!\in\! \colinvi{j},
 \text{ such that }
 \dim(\c_1 \!\cap\! V)
\!-\!\dim(\c_2 \!\cap\! V) \!\geq\! i
\big\}\\
&\!=\!
\min\left\{
\dim V : V \in \colinv, \dim(\c_1 \cap V)
-\dim(\c_2 \cap V) \geq i
\right\}\\
&\!=\! M_{R,i}(\c_1,\c_2).
\end{align*}
From \Thm{thm:monotonerdip}, we have
$\left\{ j : 
K_{R,j}(\c_1,\c_2)
 = i
\right\}
\cap
\left\{ j : 
K_{R,j}(\c_1,\c_2)
 \geq i+1
\right\}
=\emptyset$.
We thus have
\begin{align*}
M_{R,i}(\c_1,\c_2) &=
\min\left\{
j : K_{R,j}(\c_1,\c_2)
 \geq i
\right\}\\
&=
\min\left\{
j : K_{R,j}(\c_1,\c_2)= i
\right\}.
\end{align*}
Therefore the RGRW is strictly increasing with $i$ and thus
\begin{align*}
&M_{R,i}(\c_1,\c_2)
\\
&=
\min
\left\{
\dim V : V \in \colinv, \dim(\c_1 \cap V)
-\dim(\c_2 \cap V) = i
\right\},
\end{align*}
is established.
\end{IEEEproof}

Next, we show the
relation between the rank distance \cite{Gabidulin1985} and the RGRW.
Let $\phi_m:\F_{q^m}\rightarrow \F_q^{m \times 1}$ be
an $\F_q$-linear isomorphism
that expands an element of $\F_{q^m}$ as a column vector over $\F_q$
 with respect to some fixed basis for $\F_{q^m}$ over $\F_q$.
Then, we define the \textit{rank over $\F_q$} of a vector
$\vec{x}=[x_1,\dots,x_n] \in \F_{q^m}^n$,
denoted by $\mathsf{rank}_{\F_q} (\vec{x})$, as the rank of $m \times n$
matrix $\left[\phi_m(x_1), \dots, \phi_m(x_n) \right]$ over $\F_q$.
The rank distance \cite{Gabidulin1985} between two vectors
$\vec{x},\vec{y}\in\F_{q^m}^n$ is given by
$d_R(\vec{x},\vec{y}) \triangleq \rank_{\F_q}(\vec{y}-\vec{x})$.
The minimum rank distance \cite{Gabidulin1985} of a code
 $\c$ is given as
$d_R(\c)
\!\triangleq\!
\min\{d_R(\vec{x},\vec{y}): \vec{x},\vec{y} \!\in\! \c,\vec{x} \!\neq\! \vec{y}\}
\!=\! \min\{d_R(\vec{x},\vec{0}): \vec{x} \!\in\! \c, \vec{x} \!\neq\! \vec{0}\}$.
For a subspace $V\subseteq\F_{q^m}^n$,
we define by $V^* \triangleq \sum_{i=0}^{m-1} V^{q^i}$ the sum of
 subspaces $V,V^q,\dots,V^{q^{m-1}}$.

\begin{lemma}\label{lma:xxxx}
For a subspace $V \subseteq \F_{q^m}^n$ with $\dim V = 1$,
we have $\dim V^* = d_R(V)$.
\end{lemma}
\begin{IEEEproof}
Let $\vec{b} \!=\! [b_1,\dots,b_n] \!\in\! V$ be a nonzero 
vector, which implies
$\rank_{\F_q}(\vec{b})\!=\!d_R(V)$.
Let $M \!\triangleq\! \left[a_{i,j}\right]_{i,j=1}^{m,n} \!\in\! \F_{q^m}^{m \times n}$, 
$a_{i,j}\!=\!b_j^{q^{i-1}}$.
Each vector in $V^*$ is represented by
an $\F_{q^m}$-linear combination of $\vec{b},\vec{b}^q,\dots,\vec{b}^{q^{m-1}}$,
and hence $\dim V^* \!=\! \rank M$.

For $\alpha_1,\alpha_2\in\F_q$, $\beta_1,\beta_2\in\F_{q^m}$,
we have $\alpha_1 \phi_m(\beta_1) + \alpha_2 \phi_m(\beta_2) \!=\! \phi_m(\alpha_1\beta_1+\alpha_2\beta_2)$.
This implies that there always exists some $P\!\in\!\F_{q}^{n \times n}$ with $\rank P\!=\!n$
 satisfying
\begin{align}
\vec{b} P \!=\! [g_1,\dots,g_{d_R(V)},0,\dots,0] \!\in\! \F_{q^m}^n, g_j
 \!\neq\! 0, \label{eq:transform}
\end{align}
 where $g_1,\dots,g_{d_R(V)}$ are linearly independent over $\F_q$,
and note that $P$ represents the elementary column operation on $[\phi_m(b_1),\dots,\phi_m(b_n)]$.
Also for $\alpha_1,\alpha_2 \!\in\!\F_q$, $\beta_1,\beta_2\!\in\!\F_{q^m}$,
we have $\alpha_1\beta_1^{q^i} \!+\! \alpha_2\beta_2^{q^i}\!=\!(\alpha_1\beta_1 \!+\! \alpha_2\beta_2)^{q^i}$
 ($0 \!\leq\! i \!\leq\! m-1$).
Hence, for $P\!\in\!\F_{q}^{n \times n}$ satisfying \Eq{eq:transform},
we also have
$\vec{b}^{q^i} P \!=\! [g_1^{q^i},\dots,g_{d_R(V)}^{q^i},0,\dots,0] \!\in\! \F_{q^m}^n$
for all $0 \leq i \leq m-1$.
Thus, by the elementary column operation on $M$ over $\F_q$, represented
 by $P$, we get $MP$. By eliminating zero columns
from $MP$, we obtain a matrix
$M' = \left[ f_{i,j}\right]_{i,j=1}^{m,d_R(V)}$,
 $f_{i,j}=g_j^{q^{i-1}}$,
where $\rank M'=\rank M$.
Let $M'_k\in\F_{q^m}^{k \times d_R(V)}$  $(1\leq k\leq d_R(V))$ be the
 submatrix consisting of the first $k$ rows of $M'$.
Since $d_R(V) \!\leq\! \min\{m,n\}$ and $g_1,\dots,g_{d_R(V)}$ are
 linearly independent,
$M'_k$ is the generator matrix of $[d_R(V),k]$ Gabidulin code and
 $\rank M'_k=k$ \cite{Gabidulin1985}.
Thus, $M'_{d_R(V)}$ is nonsingular, and hence
we have $\rank M'_{d_R(V)}=\rank M'\!=\!d_R(V)$.
Therefore, $\dim V^* \!=\! \rank M \!=\! \rank M' \!=\!  d_R(V)$.
\end{IEEEproof}

\begin{lemma}\label{lma:rankdistance}
For a code $\c_1 \subseteq \F_{q^m}^n$ and its subcode $\c_2 \subsetneqq \c_1$,
the first RGRW can be represented as
$M_{R,1}(\c_1,\c_2) 
= \min \left\{
d_R(\vec{x},\vec{0}) : \vec{x} \in \c_1\backslash\c_2 
\right\}$.
\end{lemma}
\begin{IEEEproof}
$M_R(\c_1,\c_2)$ can be represented as
\begin{align}
&M_{R,1}(\c_1,\c_2)\nonumber\\
&=
\min \left\{
\dim W : W \!\in\! \colinv,
\dim (\c_1 \cap W)\! -\! \dim (\c_2 \cap W)\!\geq\! 1
\right\}\nonumber\\
&=
\min \Big\{
\dim W : W \in \colinv,\nonumber\\
& \exists V \!\subseteq\! W \text{\,such that\,}
V \!\subseteq\! (\c_1 \cap W), V \!\nsubseteq\! (\c_2 \cap W), \dim V \!\geq\! 1
\Big\}\label{eq:rankdistance1}.
\end{align}
For any subspace $V \subseteq \F_{q^m}^n$ with $\dim V \!\geq\! 1$,
there always exists some $W \!\in\! \colinv$ satisfying $W \!\supseteq\! V$,
because we have $V^\ast \!\in\! \colinv$ and $V^* \!\supseteq\! V$.
Also, for subspaces $W$ and $V \!\subseteq\! W$ with $\dim V \!\geq\! 1$,
 if $W$ is the smallest space in $\colinv$ including $V$, 
then $W\!=\!V^*$
 \cite{Stichtenoth1990}.
Thus \Eq{eq:rankdistance1} can be rewritten as
\begin{align}
&\min \Big\{
\dim W : V \!\subseteq\! \F_{q^m}^n, \dim V \!\geq\! 1\nonumber\\
&\quad \exists W \!\supseteq\! V, W\!\in\!\colinv, \text{\,such that\,}
V \!\subseteq\! (\c_1 \cap W), V \!\nsubseteq\! (\c_2 \cap W)
\Big\}\nonumber\\
&=\min \left\{
\dim V^* : V \!\subseteq\!\F_{q^m}^n,
 V \!\subseteq\! (\c_1 \!\cap\! V^*), V \!\nsubseteq\! (\c_2 \!\cap\! V^*),
\dim V \!\geq\! 1
\right\}\nonumber\\
&=
\min \left\{
\dim V^* : V \subseteq \c_1, V \nsubseteq \c_2,
\dim V \geq 1
\right\},\label{eq:midrankdistance}
\end{align}
where the last equality of \Eq{eq:midrankdistance} is obtained by
$V\subseteq(\c_1\cap V^*) \Leftrightarrow V\subseteq\c_1$, and
$V\nsubseteq(\c_2\cap V^*) \Leftrightarrow V\nsubseteq\c_1$ from
$V^*\supseteq V$.
For subspaces $V$ and $V' \supseteq V$,
we have $\dim V^* \leq \dim V'^*$.
Therefore, \Eq{eq:midrankdistance} can be rewritten as follows.
\begin{align*}
&\min \left\{
\dim V^* : V \subseteq \c_1, V \nsubseteq \c_2,
\dim V \geq 1
\right\}\\
&=\min \left\{
\dim V^* : V \subseteq \c_1, V \nsubseteq \c_2,
\dim V = 1
\right\}\\
&=
\min \left\{
d_R(V) : V \subseteq \c_1, V \nsubseteq \c_2,
\dim V = 1
\right\}\ \text{(by \Lma{lma:xxxx})}\\
&=
\min \left\{
d_R(\vec{x},\vec{0}) : \vec{x} \in \c_1\backslash \c_2
\right\}.
\end{align*}\\[-6.3ex]
\end{IEEEproof}

\Lma{lma:rankdistance} immediately yields the following corollary.
\begin{corollary}\label{coro:rankdistance}
For a linear code $\c$, $d_{R}(\c) = M_{R,1}(\c,\{\vec{0}\})$ holds.
\end{corollary}
This shows that $M_{R,1}(\cdot,\{\vec{0}\})$ is a generalization of
$d_R(\cdot)$.
Now we present the following proposition that generalizes the
Singleton-type bound of the rank distance \cite{Gabidulin1985}.
\begin{proposition}[Generalization of Singleton-Type Bound]\label{prop:generalizedsingleton}
Let $\c_1 \subseteq \F_{q^m}^n$ be a linear code and $\c_2 \subsetneqq \c_1$
 be its subcode.
Then, the RGRW of $\c_1$ and $\c_2$ is upper bounded by
\begin{align}
M_{R,i}(\c_1,\c_2)
\leq \min\left\{1, \frac{m}{(n-\dim \c_2)}\right\}(n-\dim \c_1)+i, \label{eq:singleton}
\end{align}
for $1 \leq i \leq \dim(\c_1/\c_2)$.
\end{proposition}
\begin{IEEEproof}
We can consider that $\c_2$ is a systematic code without loss
 of generality.
 That is, the first $\dim \c_2$ coordinates of each basis of $\c_2$ is
 one of canonical bases of $\F_{q^m}^{\dim \c_2}$.
Let $\mathcal{S}\subsetneqq\F_{q^m}^n$ be a linear code such that $\c_1$ is a
 direct sum of $\c_2$ and $\mathcal{S}$.
Then, after suitable permutation of coordinates,
a basis of $\mathcal{S}$ can be chosen such that its first $\dim \c_2$ coordinates are zero.
Then, the effective length \cite{Forney1994} of a code
 $\mathcal{S}$ is less than or equal to $n-\dim \c_2$.
Hence we have
\begin{align}
d_R(\mathcal{S})
&\leq
\min\left\{1,\frac{m}{n-\dim \c_2}\right\}(n-\dim \c_2 - \dim \mathcal{S})+1,\nonumber\\
&=
\min\left\{1,\frac{m}{n-\dim \c_2}\right\}(n-\dim \c_1)+1,\label{eq:singletonproof}
\end{align}
from the Singleton-type bound for rank metric \cite{Gabidulin1985}.

Here we write $\kappa=\min\left\{1,m/(n-\dim \c_2)\right\}$
for the sake of simplicity.
Recall that $d_R(\mathcal{S})=M_{R,1}(\mathcal{S},\{\vec{0}\})$
from \Coro{coro:rankdistance},
and $M_{R,1}(\mathcal{S},\{\vec{0}\}) \leq \kappa(n-\dim\c_1)+1$ holds
 from \Eq{eq:singletonproof}.

We shall use the mathematical induction on $t$.
We see that \Eq{eq:midproofsingleton} is true for $t=1$.
Assume that for some $t \geq 1$,
\begin{align}
M_{R,t}(\mathcal{S},\{\vec{0}\}) \leq \kappa(n-\dim\c_1) + t,
\label{eq:midproofsingleton}
\end{align}
is true.
Then, by the monotonicity shown in \Prop{lma:rgrw},
\begin{align*}
M_{R,t +1}(\mathcal{S},\{\vec{0}\})
&\leq M_{R,t}(\mathcal{S},\{\vec{0}\}) +1
\leq \kappa(n-\dim\c_1)+t +1,
\end{align*}
holds.
Thus, it is proved by mathematical induction that
\Eq{eq:midproofsingleton} holds for $1 \leq t \leq \dim (\c_1/\c_2)$.

Lastly, we prove \Eq{eq:singleton} by the above discussion about the RGRW
 of $\mathcal{S}$ and $\{\vec{0}\}$.
For an arbitrary fixed subspace $V \subseteq \F_{q^m}^n$,
we have
$\dim (\c_1 \cap V) 
\geq \dim (\mathcal{S} \cap V) + \dim (\c_2 \cap V)$,
because $\c_1$ is a direct sum of $\mathcal{S}$ and $\c_2$.
Hence,
$\dim (\c_1 \cap V) -  \dim (\c_2 \cap V)\geq \dim (\mathcal{S} \cap V)$
holds, and 
we have $M_{R,i}(\c_1,\c_2) \leq M_{R,i}(\mathcal{S},\{\vec{0}\})$
 for $1 \leq i \leq \dim (\c_1/\c_2)$ from \Def{def:rgrw}.
Therefore, from the foregoing proof,
we have 
\begin{align*}
M_{R,i}(\c_1,\c_2) \leq M_{R,i}(\mathcal{S},\{\vec{0}\})
\leq \kappa(n-\dim\c_1)+i,
\end{align*}
for $1 \leq i \leq \dim (\c_1/\c_2)$, and the proposition is proved.
\end{IEEEproof}

\Prop{prop:generalizedsingleton} immediately yields the following
corollary.
\begin{corollary}\label{coro:mrdrgrw}
For a linear code $\c \subseteq \F_{q^m}^n$,
$M_{R,i}(\c,\{\vec{0}\}) \leq \min\{1,m/n\}(n-\dim \c)+i$
 for $1 \leq i \leq \dim \c$.
The equality holds for all $i$ if and only if $\c$ is an MRD code.
\end{corollary}

\section{Universal Security Performance on Wiretap Networks}\label{sect:universalsecure}
In this section,
we express $\uequivocation_\mu$ and $\ustrong$ given in
\Sect{sect:securityperformance} in terms of the RDIP and RGRW.
From now on, we use the following definition.
\begin{definition}
For $B\!\in\!\F_{q}^{\mu \times n}$,
we define $V_B\!\triangleq\!\{\vec{u}B : \vec{u}\!\in\!\F_{q^m}^\mu\} \!\subseteq\! \F_{q^m}^n$.
\end{definition}
Recall that if an  $\F_{q^m}$-linear space $V \subseteq \F_{q^m}^n$
admits a basis in $\F_{q}^n$ then $V \in \colinv$ \cite{Stichtenoth1990},
which implies
\begin{equation}
V_B \in \colinv. \label{eq:vb}
\end{equation}

First, we give the following theorem for the universal equivocation
$\uequivocation_\mu$ given in \Def{def:universalperformance}
\begin{theorem}\label{thm:equivocation}
Consider the nested coset coding in \Def{def:nestedcoding}.
Then, the universal equivocation $\uequivocation_\mu$ of $\c_1,\c_2$ is
 given by
\begin{align*}
\uequivocation_\mu
 &= l-K_{R,\mu}(\dc_2,\dc_1).
\end{align*}
\end{theorem}
\begin{IEEEproof}
Let $B\in\F_{q}^{\mu \times n}$ be an arbitrary matrix.
By the chain rule \cite{Cover2006}, we have the following equation for the
 conditional entropy of $S$ given $BX^{\rm T}$:
\begin{align}
H(S|BX^{\rm T})
&=
H(S,X|BX^{\rm T}) - H(X|S, BX^{\rm T})
\nonumber\\
&=
H(X|BX^{\rm T}) + H(S | X, BX^{\rm T}) -
 H(X|S, BX^{\rm T})
\nonumber\\
&=
H(X|BX^{\rm T}) - H(X|S, BX^{\rm T}).
\label{eq:nonuniforms}
\end{align}
Then, from \cite[Proof of Lemma 4.2]{Zhang2009},
we have
\begin{align*}
H(X|BX^{\rm T})
&=
n-\dim \dc_1 - \dim V_B + \dim(\dc_1 \cap V_B),
\\
H(X|S, BX^{\rm T})
&=
n-\dim \dc_2 - \dim V_B + \dim(\dc_2 \cap V_B).
\end{align*}
By substituting these equations into \Eq{eq:nonuniforms},
we have
\begin{align}
H(S|BX^{\rm T})
&=
 \dim \dc_2 \!-\! \dim \dc_1 \!-\! \dim(\dc_2 \cap V_B)
 \!+\! \dim(\dc_1 \cap V_B) \nonumber\\
&=
l - \dim(\dc_2 \cap V_B)
 +
 \dim(\dc_1 \cap V_B).\label{eq:nonuniforms2}
\end{align}
By \Eq{eq:vb} we have
\begin{align}
\left\{ V_B : B\in \F_{q}^{\mu\times n}\right\} = \bigcup_{i\leq \mu} \colinvi{i}.
\label{eq:m3}
\end{align}
Thus, by \Eq{eq:nonuniforms2} and \Def{def:rdip},
the universal equivocation $\uequivocation_\mu$  is given as follows.
\begin{align*}
&\uequivocation_\mu
 =
 \min_{B \in \F_{q}^{\mu\times n}}
 H(S|BX^{\rm T})\\
&= l- \max_{B\in \F_{q}^{\mu\times n}}
\left\{
\dim(\dc_2 \cap V_B) - \dim(\dc_1 \cap V_B)
\right\}\\
 &=
 l-
 \max_{V \in \bigcup_{i\leq \mu} \colinvi{i}}
\left\{
\dim(\dc_2 \cap V) - \dim(\dc_1 \cap V)
\right\}\mbox{(by \Eq{eq:m3})}\\
 &=
 l-
 \max_{V \in \colinvi{\mu}}
\left\{
\dim(\dc_2 \cap V) - \dim(\dc_1 \cap V)
\right\}\mbox{(by Thm.\  \ref{thm:monotonerdip})}\\
 &=
 l-K_{R,\mu}(\dc_2,\dc_1).
\end{align*}\\[-6.3ex]
\end{IEEEproof}

\begin{example}\label{ex1}
The existing schemes \cite{Kurihara2012,Silva2009,Silva2011}
used MRD codes as 
$\dc_1$ and $\dc_2$, where $m \geq n$.
By \Coro{coro:rankdistance},
we have $\dim (V \cap \dc_2) = 0$
for any $V \in \colinvi{\dim \c_2}$.
This implies $K_{R,\mu}(\dc_2, \dc_1)=K_{R,\mu}(\dc_2, \{\vec{0}\})=0$
for $0\leq \mu \leq \dim \c_2$.

On the other hand, $K_{R,\dim \c_1}(\dc_2, \{\vec{0}\}) = \dim \c_1 - \dim \c_2$ by \Coro{coro:mrdrgrw}. 
Since $\dim (V \cap \dc_1) \!=\! 0$
for any $V \!\in\! \colinvi{\dim \c_1}$ by \Coro{coro:rankdistance},
we have $K_{R,\dim \c_1}(\dc_2, \dc_1) \!=\! \dim \c_1 \!-\! \dim \c_2$.
By \Thm{thm:monotonerdip},
$K_{R,\mu}(\dc_2$, $\dc_1) \!=\! \mu \!-\! \dim \c_2$
for $\dim \c_2 \!\leq\! \mu \!\leq\! \dim \c_1$.

By \Thm{thm:equivocation}, we see that
$\uequivocation_\mu \!=\! l \!-\! \max\{0, \mu \!-\! \dim\c_2\}$
for $0 \!\leq\! \mu \!\leq\! \dim \c_1 (= l \!+\! \dim \c_2)$ in
the  schemes \cite{Kurihara2012,Silva2009,Silva2011}.
\end{example}

We then have the following corollary by the RGRW.
\Coro{prop:perfectsecrecy} shows that the wiretapper obtain no
information of $S$ from any $M_{R,1}(\dc_2,\dc_1)-1$ links.
\begin{corollary}\label{prop:perfectsecrecy}
Consider the nested coset coding in \Def{def:nestedcoding}.
Then, the wiretapper must observe at least $M_{R,j}(\dc_2,\dc_1)$ links to
 obtain the mutual information $j$ ($1 \leq j \leq l$) between $S$
 and observed packets.
\end{corollary}
\begin{IEEEproof}
From \Eq{eq:nonuniforms2},
the smallest number $\mu$ of tapped links satisfying
 $I(S;BX^{\rm T})=j$ $(1 \leq j \leq l)$
 is
\begin{align*}
&\min \left\{ \mu: \exists B \in \F_q^{\mu \times n}, I(S;BX^{\rm T})=j\right\} \\
&\ =
\min \left\{\mu: \exists B \in \F_q^{\mu \times n}, l-H(S|BX^{\rm T})=j\right\}\\
&\ =
\min \left\{\mu : \exists B \in \F_q^{\mu \times n}, \dim (\dc_2 \cap V_B) - \dim(\dc_1 \cap V_B)=j\right\}.
\end{align*}
From \cite[Lemma 1]{Stichtenoth1990} and \Lma{lma:rgrw},
this equation can be rewritten as follows.
\begin{align*}
&\min
\left\{\mu : \exists B \in \F_q^{\mu \times n}, \dim (\dc_2 \cap V_B) - \dim(\dc_1 \cap V_B)=j\right\}
\\
&=
\min \left\{\dim V : V \in \colinv, \dim (\dc_2 \cap V) - \dim(\dc_1 \cap V)=j\right\}
\\
&= M_{R,j}(\dc_2,\dc_1).
\end{align*}\\[-6.3ex]
\end{IEEEproof}

Although the message $S$ has
 been assumed to be uniformly distributed over $\F_{q^m}^l$ in
 \Sect{sect:nestedcoding},
the following proposition reveals that
the wiretapper still obtain no information of $S$ from any
$M_{R,1}(\dc_2,\dc_1)-1$ links even if $S$ is arbitrarily distributed.

\begin{proposition}\label{coro:distribution}
Fix the transfer matrix $B$ to the wiretapper.
Suppose that the wiretapper obtain no information of $S$ from $BX^{\rm T}$
when $S$ is uniformly distributed over $\F_{q^m}^l$ as described in
 \Sect{sect:nestedcoding}.
Then, even if $S$ is chosen according to an arbitrary distribution
 over $\F_{q^m}^l$,
 the wiretapper still obtain no information of $S$ from $BX^{\rm T}$, that is, $I(S;BX^{\rm T})=0$.
\end{proposition}
\begin{IEEEproof}
When we assume that $S$ is arbitrarily distributed over
 $\F_{q^m}^l$,
$H(X|S, BX^{\rm T})$ is
upper bounded as follows from
 \cite[Proof of Lemma 6]{Silva2011} and \cite[Proof of Lemma 4.2]{Zhang2009}.
\begin{align*}
H(X|S, BX^{\rm T})
&\leq
n-\dim\dc_2-\dim V_B + \dim(\dc_2\cap V_B).
\end{align*}
Also, since $X$ is uniformly distributed over
 a coset $\psi(S)\in\c_1/\c_2$ for fixed $S$,
we have $H(X|S)=\dim \c_2=n-\dim\dc_2$.
For the dimension of a subspace $\{B X^{\rm T} : X\in\c_1\}$,
we have
\begin{align*}
&\dim \{B X^{\rm T} : X \in \c_1\}
=\rank B G^{\rm T}
=\rank G B^{\rm T}\\
&\quad=\dim \{G \vec{v}^{\rm T} : \vec{v} \in V_B\}
=\dim V_B - \dim (\dc_1 \cap V_B),
\end{align*}
where $G\in\F_{q^m}^{\dim \c_1 \times n}$ is a generator matrix of $\c_1$.
Hence we have
$H(BX^{\rm T})\leq  \dim V_B-\dim (\dc_1 \cap V_B)$.
We thus have
\begin{align}
I(S;BX^{\rm T})
&=I(S,X;BX^{\rm T}) -
 I(X;BX^{\rm T}|S)\nonumber\\
&=H(BX^{\rm T}) - H(X|S) + H(X|S,
 BX^{\rm T})\nonumber\\
&\leq
\dim(\dc_2\cap V_B) - \dim(\dc_1\cap V_B)\label{eq:m1}
\end{align}
for any distribution of $S$.
By $I(S;BX^{\rm T})=H(S)-H(S|BX^{\rm T})$ and \Eq{eq:nonuniforms2} we can see that
 the equality holds if $S$ is uniformly distributed.
Therefore, for fixed $B$, if $I(S;BX^{\rm T})=0$ holds for
 uniformly distributed $S$,
then the right hand side of \Eq{eq:m1} is zero,
which implies that $I(S;BX^{\rm T})=0$ also holds for arbitrarily
 distributed $S$ from the nonnegativity of mutual information \cite{Cover2006}.
\end{IEEEproof}

Lastly, we express $\ustrong$ in
\Def{def:universalalpha} in terms of the RGRW.
For a subset $\mathcal{J} \subseteq \{1,\dots,N\}$ and a vector $\vec{c}=[c_1,\dots,c_N]\in\F_{q^m}^N$,
let $P_\mathcal{J}(\vec{c})$ be a vector of length $|\mathcal{J}|$ over $\F_{q^m}$,
obtained by removing the $t$-th components $c_t$ for $t \notin \mathcal{J}$.
For example for $\mathcal{J}=\{1,3\}$ and $\vec{c}=[1,1,0,1]$ ($N=4$),
we have $P_\mathcal{J}(\vec{c})=[1,0]$.
The \textit{punctured code} $P_\mathcal{J}(\c)$ of a
code $\c \in \F_{q^n}^N$ is given by
$P_\mathcal{J}(\c)
\triangleq
\left\{P_\mathcal{J}(\vec{c})
 : \vec{c}\in \c \right\}$.
The \textit{shortened code} $\c_\mathcal{J}$ of a code $\c \subseteq \F_{q^m}^N$
is defined by
$\c_\mathcal{J} \triangleq \left\{
P_\mathcal{J}(\vec{c})
: \vec{c}=[c_1,\dots,c_N] \in \c, c_i = 0 \text{ for } i \notin \mathcal{J}
\right\}$.
For example for $\c=\{[0,0,0],[1,1,0],[1,0,1],[0,1,1]\}$ ($N=3$) and
$\mathcal{J}=\{2,3\}$,
we have $\c_\mathcal{J}=\{[0,0],[1,1]\}$.
We then have the following theorem for
the universal $\ustrong$-strong security defined in \Def{def:universalalpha}.

\begin{theorem}\label{thm:universalalpha}
Let $\bari{i}\triangleq\{1,\dots,l+n\}\backslash\{i\}$.
Fix $\c_1$, $\c_2$ and $\psi$ in \Def{def:nestedcoding}
and consider the corresponding
nested coset coding scheme in \Def{def:nestedcoding}.
By using $\c_1$, $\c_2$ and $\psi$, define
\begin{align*}
\c'_1 \triangleq \left\{[S,X]: S \in \F_{q^m}^l\text{
 and } X\in\psi(S)\right\} \subseteq \F_{q^m}^{l+n}.
\end{align*}
For each index $1 \leq i \leq l$,
we define a punctured code $\d_{1,i}$ of $\c'_1$ as
 $\d_{1,i} \triangleq P_{\bari{i}}(\c'_1) \subseteq\F_{q^m}^{l+n-1}$,
and a shortened code $\d_{2,i}$ of $\c'_1$ as
 $\d_{2,i} \triangleq (\c'_1)_{\bari{i}} \subseteq\F_{q^m}^{l+n-1}$.
Then, the value $\ustrong$ in \Def{def:universalalpha}
 is given by
\begin{align}
\ustrong = \min
\left\{
M_{R,1}(\dd_{2,i},\dd_{1,i}) : 1 \leq i \leq l
\right\} -1.
\label{eq:universalalphamin}
\end{align}
\end{theorem}
\begin{IEEEproof}
Define $\c'_2 \triangleq \{[\vec{0},\vec{c}_2] : \vec{c}_2 \in \c_2\}\subseteq\F_{q^m}^{l+n}$.
Since $\c_2 \subsetneqq \c_1$, $\c'_2$ is also a subcode of $\c'_1$.
Thus, in terms of $\c'_1$ and $\c'_2$, we can see that the vector
 $[S,X]\in\F_{q^m}^{l+n}$ is generated by a nested coset
 coding scheme of $\c'_1$ and $\c'_2$ from $S$.
Then, from the definition of $\c'_1$ and $\c'_2$,
we can see that $\d_{2,i}$ is a subcode of $\d_{1,i}$ with dimension
 $\dim \d_{2,i}=\dim \d_{1,i} - 1 = \dim \c_1 - 1$ over $\F_{q^m}$
for each $i \in \{1,\dots,l\}$.

Let $\mathcal{L}\triangleq\{1,\dots,l\}$
and $S_{\mathcal{L}\backslash\{i\}} \triangleq [S_1,\dots,S_{i-1},S_{i+1},\dots,S_l]$ for
 each $1\leq i \leq l$.
For $S_i \in\F_{q^m}$ define a coset
\begin{align*}
\phi(S_i)
&\triangleq \left\{[S_{\mathcal{L}\backslash\{i\}},X]:
 S_{\mathcal{L}\backslash\{i\}} \in\F_{q^m}^{l-1}
 \text{ and }
 X \in \psi(S)\right\}
\in\d_{1,i}/\d_{2,i}.
\end{align*}
Here we define
$\vecxi \triangleq P_{\bari{i}}([S,X])= [S_{\mathcal{L}\backslash\{i\}},X] \in \d_{1,i}$.
Recall that $S_1,\dots,S_l$ are mutually independent and
 uniformly distributed over $\F_{q^m}$.
Thus, considering a nested coset coding scheme that generates
$\vecxi$ from a secret message $S_i \in \F_{q^m}$ with $\d_1,\d_2$,
we can see that
$\vecxi \in \phi(S_i) \in \d_{1,i}/\d_{2,i}$
 is chosen uniformly at random from $\phi(S_i)$.
Therefore, 
we have $I(S_i;D\vecxi^{\rm T})=0$
for any $D\in\F_q^{\mu \times (n+l-1)}$ whenever $\mu < M_{R,1}(\dd_{2,i},\dd_{1,i})$
from \Coro{prop:perfectsecrecy}.

For an arbitrary subset $\mathcal{R}\!\subseteq\!\mathcal{L}\backslash\{i\}$,
define a matrix $F_\mathcal{R}$ that consists of $|\mathcal{R}|$ rows of
 an $(l-1)\times(l-1)$ identity matrix, satisfying
$[S_j : j \!\in\! \mathcal{R}]^{\rm T} \!=\! F_\mathcal{R} S_{\mathcal{L}\backslash\{i\}}^{\rm T}$.
For an arbitrary matrix $B\!\in\!\F_q^{k \times n}$ ($0 \!\leq\! k \!\leq\! n$),
set $D\!=\!\left[\begin{smallmatrix}F_\mathcal{R} & O \\ O & B\end{smallmatrix}\right]$. 
Then, from the foregoing proof, we have
\begin{align*}
0=I(S_i;D\vecxi^{\rm T})&=
I(S_i; S_\mathcal{R},BX^{\rm T})
=H(S_i|S_\mathcal{R}) - H(S_i | BX^{\rm T},S_{\mathcal{R}})
\\
&=
H(S_i) - H(S_i | BX^{\rm T},S_{\mathcal{R}})
=I(S_i; BX^{\rm T}|S_\mathcal{R}),
\end{align*}
whenever $|\mathcal{R}| \!+\! k \!<\! M_1(\dd_{2,i},\dd_{1,i})$.
Since $I(S_i; BX^{\rm T}|S_\mathcal{R})\!=\!0$
 is equivalent to \Eq{eq:universalalpha} from \cite[Prop.\@\,5]{Silva2009},
we have \Eq{eq:universalalphamin} by selecting the minimum value of
 $M_{R,1}(\dd_{2,i},\dd_{1,i})\!-\! 1$ for $1 \!\leq\! i \!\leq\! l$.
\end{IEEEproof}

\begin{example}\label{ex2}
The scheme proposed in \cite{Kurihara2012}
used a systematic MRD code as $\c'_1$ (not $\c_1$), where $m \geq l+n$.
We proved 
$\min
\left\{
M_{R,1}(\dd_{2,i},\dd_{1,i}) : 1 \leq i \leq l
\right\}=n$ in \cite[Proof of Theorem 4]{Kurihara2012}.
By \Thm{thm:universalalpha}, we see
that the scheme \cite{Kurihara2012} attains 
the universal $(n-1)$-strong security in the sense of
\Def{def:universalalpha}, while
\cite{Kurihara2012} proved it by adapting the
proof argument in \cite{Silva2009}.
\end{example}

As shown in \Prop{coro:distribution},
no information of $S$ is leaked from less than
$M_{R,1}(\dc_2,\dc_1)$ tapped links
even if $S$ is arbitrarily distributed.
In contrast, $S$ must be uniformly distributed over $\F_{q^m}^l$
to establish \Thm{thm:universalalpha}.
This is because elements of $S$ need to be treated as extra random
packets, as in strongly secure network coding schemes \cite{Silva2009,Harada2008,Matsumoto2011}.

\section{Universal Error Correction Capability of Secure Network Coding}\label{sect:errorcorrection}
This section derives the universal error correction
capability by the approach of
\cite[Section III]{Silva2009a}.
Recall that the received packets $Y$ is given by
 $Y^{\rm T}=AX^{\rm T}+DZ^{\rm T}$ in the setup of \Sect{sect:nestedcoding},
and that $X$ is chosen from the coset $\psi(S)\in\c_1/\c_2$ corresponding to $S$ by
the nested coset coding in \Def{def:nestedcoding}.
From now on, we write $\mathcal{X}\triangleq\psi(S)$ for the sake of simplicity.

First, we define the \textit{discrepancy} \cite{Silva2009a} between
$\mathcal{X}$ and $Y$ by
\begin{align}
\Delta_A (\mathcal{X}, Y)
&\!\triangleq\!
\min \{r \!\in\! \mathbb{N} :
D \!\in\! \F_{q}^{N \times r},
Z \!\in\! \F_{q^m}^{r},
X \!\in\! \mathcal{X},
Y^{\rm T}\!=\!AX^{\rm T}\!+\!DZ^{\rm T}\}
\nonumber\\
&\!=\!
\min \left\{
d_R(XA^{\rm T},Y) : X \in \mathcal{X}
\right\},\label{eq:defdescrepancy}
\end{align}
where the second equality is derived from \cite[Lemma 4]{Silva2009a}.
This definition of $\Delta_A(\mathcal{X},Y)$ represents 
the minimum number $r$
of error packets $Z$ required to be injected in order to transform
at least one element of $\mathcal{X}$ into $Y$,
as \cite[Eq.\,(9)]{Silva2009}.

Next, we define the \textit{$\Delta$-distance} \cite{Silva2009a}
between
$\mathcal{X}$ and $\mathcal{X}'$, induced by
$\Delta_A(\mathcal{X},Y)$, as
\begin{align}
\delta_A(\mathcal{X},\mathcal{X}')
\triangleq
\min
\left\{\Delta_A(\mathcal{X},Y)+ \Delta_A(\mathcal{X}',Y) : Y \in \F_{q^m}^N\right\},
\label{eq:defdelta}
\end{align}
for $\mathcal{X},\mathcal{X}'\in\c_1/\c_2$.
\begin{lemma}\label{lma:deltadistance}
For $\mathcal{X}, \mathcal{X}' \in \c_1/\c_2$, we have
\begin{align}
\delta_A(\mathcal{X},\mathcal{X}')
&=
\min
\left\{
d_R(XA^{\rm T},X'A^{\rm T}):
X \in \mathcal{X},
X' \in \mathcal{X}'
\right\}. \label{eq:deltalma}
\end{align}
\end{lemma}
\begin{IEEEproof}
First we have
\begin{align}
&\delta_A(\mathcal{X},\mathcal{X}')
=\min
\left\{\Delta_A(\mathcal{X},Y)+ \Delta_A(\mathcal{X}',Y) : Y \in \F_{q^m}^N\right\}
\nonumber\\
&\!=\!\min
\Big\{
\min \left\{
d_R(XA^{\rm T},Y) : X \in \mathcal{X}
\right\}
\nonumber\\
&\qquad\qquad+
\min \left\{
d_R(X'A^{\rm T},Y) : X' \in \mathcal{X}'
\right\}
: Y \in \F_{q^m}^N
\Big\}\nonumber\\
&\!=\!
\min \left\{
d_R(XA^{\rm \!T},Y) \!+\! d_R(X'\!A^{\rm \!T},Y):
X \!\in\! \mathcal{X},
X'\! \!\in\! \mathcal{X}'\!,
Y \!\in\! \F_{q^m}^N
\right\}.\label{eq:triangle}
\end{align}
The rank distance satisfies 
the triangle inequality
$d_R(XA^{\rm T},XA^{\rm T}) \leq d_R(XA^{\rm T},Y) + d_R(X'A^{\rm T},Y)$
for $\forall Y\in\F_{q^m}^N$ \cite{Gabidulin1985}.
This lower bound can be achieved by choosing,
\eg $Y =X A^{\rm T}$.
Therefore, from \Eq{eq:triangle}, we have \Eq{eq:deltalma}.
\end{IEEEproof}

The next lemma shows that $\Delta_A(\mathcal{X},Y)$
is \textit{normal} \cite[Definition 1]{Silva2009a}.
\begin{lemma}\label{lma:normal}
For all $\mathcal{X},\mathcal{X}'\in\c_1/\c_2$ and all $0 \leq i \leq \delta_A(\mathcal{X},\mathcal{X}')$,
there exists some $Y\in\F_{q^m}^n$ such that $\Delta_A(\mathcal{X},Y)=i$
 and $\Delta_A(\mathcal{X}',Y)=\delta_A(\mathcal{X},\mathcal{X}')-i$.
\end{lemma}
\begin{IEEEproof}
Let $\mathcal{X},\mathcal{X}'\in\c_1/\c_2$ and let
 $0 \leq i \leq d=\delta_A(\mathcal{X},\mathcal{X}')$.
Then, $d=\min \left\{ d_R(XA^{\rm T},X'A^{\rm T}): X \in \mathcal{X}, X' \in \mathcal{X}'\right\}$
from \Lma{lma:deltadistance}.
Let $\bar{X}\in\mathcal{X}$ and $\bar{X'}\in\mathcal{X}'$ be vectors
 satisfying $d=d_R(\bar{X}A^{\rm T},\bar{X}'A^{\rm T})$.
From the proof of \cite[Theorem 6]{Silva2009a},
we can always find two vectors $W,W'\in\F_{q^m}^n$ such that
$W+W' = (\bar{X}'-\bar{X})A^{\rm T}$, $\rank_{\F_q}(W)=i$ and
 $\rank_{\F_q}(W')=d-i$.
Taking $\bar{Y}=\bar{X}A^{\rm T}+W=\bar{X}'A^{\rm T}-W'$,
we have
$d_R(\bar{X}A^{\rm T},\bar{Y}) = i$ and $d_R(\bar{X}'A^{\rm T},\bar{Y}) = d-i$.
We thus obtain
$\Delta_A(\mathcal{X},\bar{Y}) \leq i$ and
$\Delta_A(\mathcal{X}',\bar{Y}) \leq d-i$ from \Eq{eq:defdescrepancy}.
On the other hand, since $\delta_A(\mathcal{X},\mathcal{X}')=d$,
we have $\Delta_A(\mathcal{X},Y) + \Delta_A(\mathcal{X}',Y) \geq d$
for any $Y \in \F_{q^m}^n$ from from \Eq{eq:defdelta}.
Therefore, 
$\Delta_A(\mathcal{X},\bar{Y}) = i$ and $\Delta_A(\mathcal{X}',\bar{Y})
 = d-i$
hold.
\end{IEEEproof}
Let $\delta_A(\c_1/\c_2)$ be the minimum $\Delta$-distance given by
\begin{align*}
\delta_A (\c_1/\c_2)
\triangleq
\min \left\{ \delta_A(\mathcal{X},\mathcal{X}') :
 \mathcal{X},\mathcal{X}' \in \c_1/\c_2, \mathcal{X}\neq\mathcal{X}'
\right\}.
\end{align*}
As \cite[Theorem 7]{Silva2009a},
from \Lma{lma:normal} and \cite[Theorem 3]{Silva2009a},
we have the following proposition.
\begin{proposition}\label{prop:deltacorrection}
A nested coset coding scheme with $\c_1,\c_2$ is guaranteed to
determine the unique coset $\mathcal{X}$ against
any $t$ packet errors for any fixed $A$
if and only if
$\delta_A(\c_1/\c_2) \!>\! 2t$. \hfill\IEEEQED
\end{proposition}
Here we note that if $\mathcal{X}$ is uniquely determined, $S$ is also uniquely
determined from \Def{def:nestedcoding}.

\begin{lemma}\label{eq:cosetdelta}
$\delta_A(\c_1/\c_2) =
\min\{
d_R(XA^{\rm\! T}, X'\!A^{\rm\! T}) : X,X' \!\!\in\! \c_1, X'\!\!-\!X \!\notin\! \c_2
\}$.
\end{lemma}
\begin{IEEEproof}
\begin{align*}
&\delta_A(\c_1/\c_2)
=\min \left\{ \delta_A(\mathcal{X},\mathcal{X}') :
 \mathcal{X},\mathcal{X}' \in \c_1/\c_2, \mathcal{X}\neq\mathcal{X}'
\right\}\\
&\!=\!
\min \!\Big\{
\!\min\!\left\{
d_R(XA^{\rm\! T},X'\!A^{\rm\! T})\!:\!X \!\in\!\mathcal{X},X'\!\!\in\!\mathcal{X}'
\right\} \!:\!
\mathcal{X},\mathcal{X}'\!\!\in\! \c_1/\c_2, \mathcal{X}\!\neq\!\mathcal{X}'
\Big\}\\
&\!=\!
\min\Big\{
d_R(XA^{\rm T},X'A^{\rm T}):
X \!\in\!\mathcal{X} \!\in\! \c_1/\c_2,
X'\!\in\!\mathcal{X}'\!\in\!\c_1/\c_2, \mathcal{X}\!\neq\!\mathcal{X}'
\Big\}\\
&\!=\!
\min \left\{
d_R(XA^{\rm T},X'A^{\rm T}) : X,X' \in \c_1, X'-X \notin \c_2
\right\}.
\end{align*}\\[-6.3ex]
\end{IEEEproof}

\begin{theorem}\label{thm:errorcorrectioncap}
Consider the nested coset coding in \Def{def:nestedcoding}.
Then, the scheme is a universally (\ie simultaneously
for all $A \in \F_q^{N \times n}$ with rank deficiency at most $\rho$)
 $t$-error-$\rho$-erasure-correcting secure network coding
 if and only if $M_{R,1}(\c_1,\c_2) > 2t + \rho$.
\end{theorem}
\begin{IEEEproof}
For the rank deficiency $\rho \!=\! n\!-\!\rank A$,
we have $d_R(X,X')\!-\!\rho \!\leq\! d_R(XA^{\rm T},X'A^{\rm T})$,
and there always exists $A \in \F_q^{N\times n}$
depending on $(X,X')$ such that the equality holds.
Thus, from \Lma{eq:cosetdelta}, we have
\begin{align*}
\min_{\substack{A \in \F_q^{N \times n}:\\ \rank A = n-\rho}}
\delta_A(\c_1/\c_2)
&\!=\!
\min \left\{
d_R(X,X') : X,X' \!\in\! \c_1, X'\!-\!X \!\notin\! \c_2
\right\}\!-\!\rho\\[-2ex]
&\!=\!
\min \left\{
d_R(X,\vec{0}) : X \in \c_1, X \notin \c_2
\right\}-\rho\\
&\!=\!
M_{R,1}(\c_1,\c_2) -\rho. \quad \text{(by \Lma{lma:rankdistance})}
\end{align*}
Therefore, we have
${\displaystyle \min_{A: \rank A = n-\rho}\!\!\delta_A(\c_1/\c_2)
\!<\! \min_{A: \rank A = n-\rho'}\!\!\delta_A(\c_1/\c_2)}$
 for $\rho > \rho'$, and hence we obtain
${\displaystyle \min_{A: \rank A \geq n-\rho}\delta_A(\c_1/\c_2)=}$
${\displaystyle \min_{A: \rank A = n-\rho}\delta_A(\c_1/\c_2)= M_{R,1}(\c_1,\c_2) \!-\!\rho}$.
\end{IEEEproof}

\begin{example}\label{ex3}
The existing scheme \cite{Silva2011} used MRD codes as $\c_1,\c_2$,
 where $m \geq n$.
Then, by \Coro{coro:mrdrgrw},
 we have $M_{R,1}(\c_1,\{\vec{0}\})=n-\dim\c_1+1$.
Since $\dim (V \cap \c_2)=0$ for any $V \in \colinvi{\dim\dc_2}$ by
 \Coro{coro:rankdistance} and $\dim \dc_2 > n-\dim \c_1$,
 we have $M_{R,1}(\c_1,\c_2)=M_{R,1}(\c_1,\{\vec{0}\})$.
Thus, by \Thm{thm:errorcorrectioncap} and \Coro{coro:rankdistance},
the scheme is universally $t$-error-$\rho$-erasure-correcting
 when $M_R(\c_1,\{\vec{0}\})=d_R(\c_1) > 2t+\rho$, as shown in \cite[Theorem 11]{Silva2011}.
\end{example}

\textsc{Acknowledgment:}
This research was partially supported by 
the MEXT  Grant-in-Aid for Scientific Research (A) No.~23246071.



\end{document}